\documentclass{ws-jai}
\usepackage[flushleft]{threeparttable}
\usepackage{xcolor}
\newsavebox{\imgbox}
\newsavebox{\blockbox}
\usepackage{subcaption}
\usepackage{xurl}
\setcitestyle{aysep={}}
\usepackage{adjustbox}
\usepackage{placeins}  
\newsavebox{\capbox}
\usepackage[verbose,hypertexnames=false]{hyperref}
\hypersetup{colorlinks=false,allbordercolors=blue,pdfborderstyle={/S/U/W 1}}

\renewcommand{\wsdraftnote}{}
\begin{document}

\markboth{}{Correlated Radio Frequency Interference Monitor for Radio Telescopes}

\title{Correlated Radio Frequency Interference Monitor for Radio Telescopes}

\author{Lindsay Berkhout$^{1,2*}$, Ian Hendricksen$^{1,2,3}$, Joshua Montgomery$^{3}$, Arnab Chakraborty$^{1,2,4}$, Jean-Francois Cliche$^{1,2}$, Matt Dobbs$^{1,2}$,  Albert Wai Kit Lau$^{4}$, James Mertens$^{4,6}$, Andre Renard$^{4}$, Keith Vanderlinde$^{4,5}$, Dallas Wulf$^{1,2}$}

\address{
$^{1}$Department of Physics, McGill University, 3600 rue University, Montreal, QC H3A 2T8, Canada\\
$^{2}$Trottier Space Institute at McGill, McGill University, 3550 rue University, Montreal, QC H3A 2A7, Canada\\
$^{3}$t0.technology, 300 rue Leo-Pariseau, Montreal, Canada\\
$^{4}$Dunlap Institute for Astronomy \& Astrophysics, University of Toronto, 50 St. George Street, Toronto, ON M5S 3H4, Canada\\
$^{5}$David A. Dunlap Department of Astronomy \& Astrophysics, University of Toronto, 50 St. George Street, Toronto, Canada\\
$^{6}$ CERCA/ISO, Department of Physics, Case Western Reserve University, 10900 Euclid Avenue, Cleveland, OH 44106, USA} 

\maketitle

\corres{$^{*}$Corresponding author libby.berkhout@mcgill.ca}

\begin{abstract}
Radio frequency interference (RFI) is one of the biggest challenges facing current radio astronomy experiments. Telescopes like the Canadian Hydrogen Observatory and Radio transient Detector (CHORD) need to accurately identify RFI to deliver on their science goals. For applications like HI intensity mapping for precision cosmology, long integrations consisting of hundreds to thousands of hours are needed, and thorough RFI flagging is required to avoid contaminating deep datasets. Currently, many major precision cosmology instruments are facing systematic limitations that include unflagged RFI contamination. For transients, false triggers due to RFI are common and interfere with the detection of real sources like fast radio bursts (FRBs). This paper presents a prototype design for an auxiliary array of horizon-pointed antennas that are cross-correlated in real-time at full cadence with the main telescope to assist in the identification and mitigation of RFI for the CHORD instrument. The major benefit of the version proposed here over a more traditional omnidirectional observatory monitor is the capability to synchronize with CHORD in real time and provide fine timescale flags and cross correlations. The general setup and techniques used here will be applicable to other radio telescopes as well. 
\end{abstract}

\section{Introduction}
Radio telescopes face an increasingly contaminated electromagnetic spectrum. Radio Frequency Interference (RFI) — any radio frequency signal that does not originate from an astronomical target — is a major source of data loss and contamination in radio astronomy experiments. Over time, the electromagnetic spectrum becomes increasingly crowded as more devices come online; recent examples include the deployment of Starlink satellite constellations (\citealt{lofarsecondgen}, \citealt{lofarstarlink}, \citealt{skastarlink}) and the rollout of 5G cellular networks and other new cellular services (\citealt{nhan2026lessonslearnedongoingcoordination}, \citealt{5Gmodels}). The Canadian Hydrogen Intensity Mapping Experiment (CHIME) reports on average 42 percent of their 400--800~MHz observing range lost to RFI (\citealt{2022chime}).

\begin{figure}[htbp!]
    \centering
    \includegraphics[width=.7\linewidth]{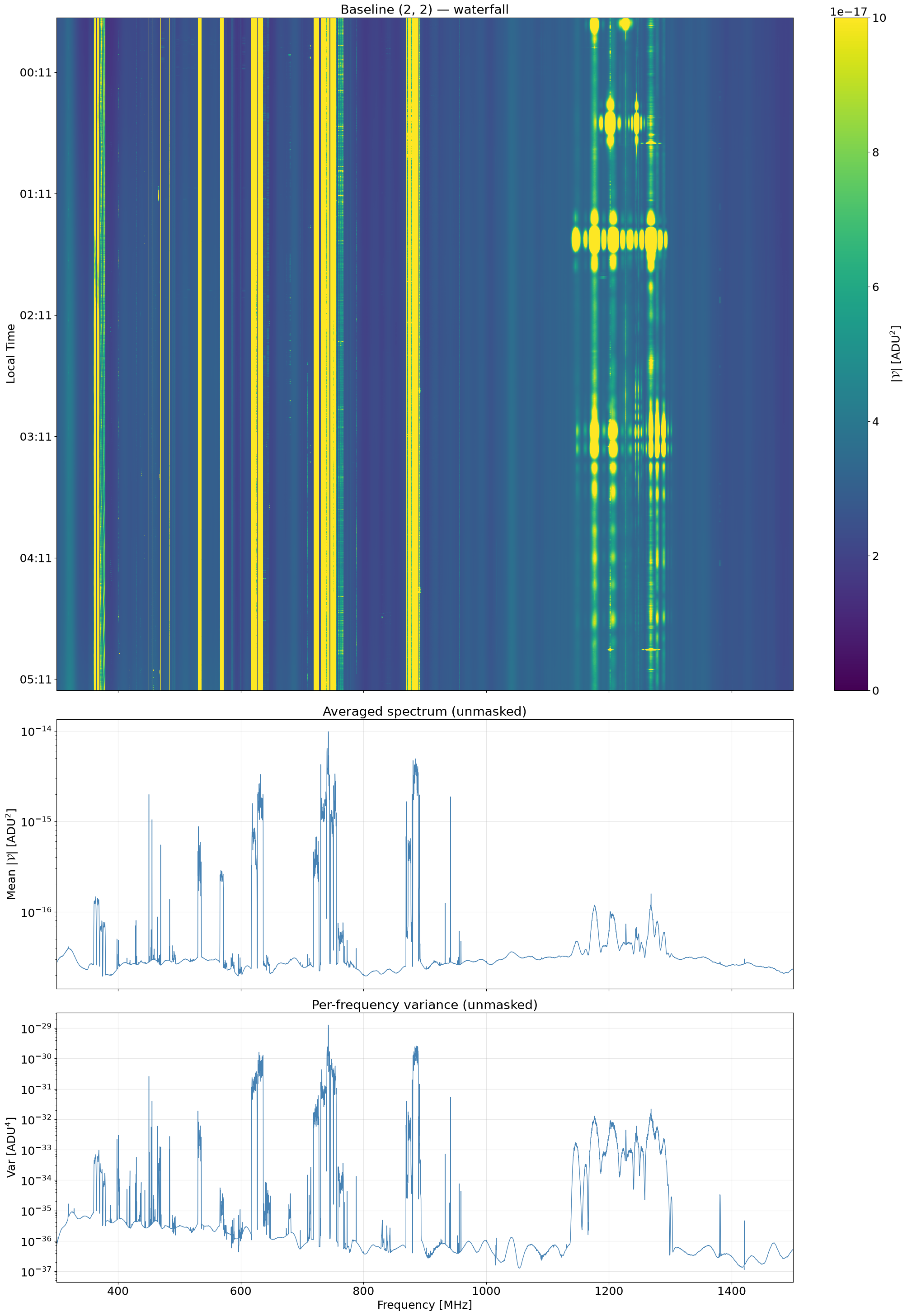}
    \caption{A typical example of the RFI environment at the Dominion Radio Astrophysical Observatory (DRAO) as seen by one CHORD dish. (Top) An approximately 5-hour autocorrelation waterfall over 300--1500~MHz from one CHORD dish, (middle) the average spectrum of the waterfall, and (bottom) the spectrum variance of the waterfall. The most prominent structure in the waterfall highlights the bright, persistent RFI, while the spectrum variance highlights the variable nature of some of the sources. Bands like GNSS that have variation due to satellite visibility also show high variance. }
    \label{fig:rfispec}
\end{figure}

This work will focus on the RFI environment at the Canadian Hydrogen Observatory and Radio transient Detector (CHORD, \citealt{chord2026}) telescope, but the techniques and instrumentation are more generally applicable to radio observatories. CHORD is co-located with CHIME at the Dominion Radio Astrophysical Observatory (DRAO) in British Columbia, Canada, with a 300--1500~MHz frequency band which partially overlaps with that of CHIME. The array will be 512 6-meter parabolic dishes with dual polarization feeds, with a 64 dish pathfinder currently being commissioned. Two outrigger telescopes with 64 dishes each are being constructed at the Hat Creek Radio Observatory (HCRO) in California and the Green Bank Observatory (GBO) in West Virginia for improved angular resolution and localization. Here we focus on the site of the CHORD main array at DRAO.

\begin{figure}
    \centering
    \includegraphics[width=0.9\linewidth]{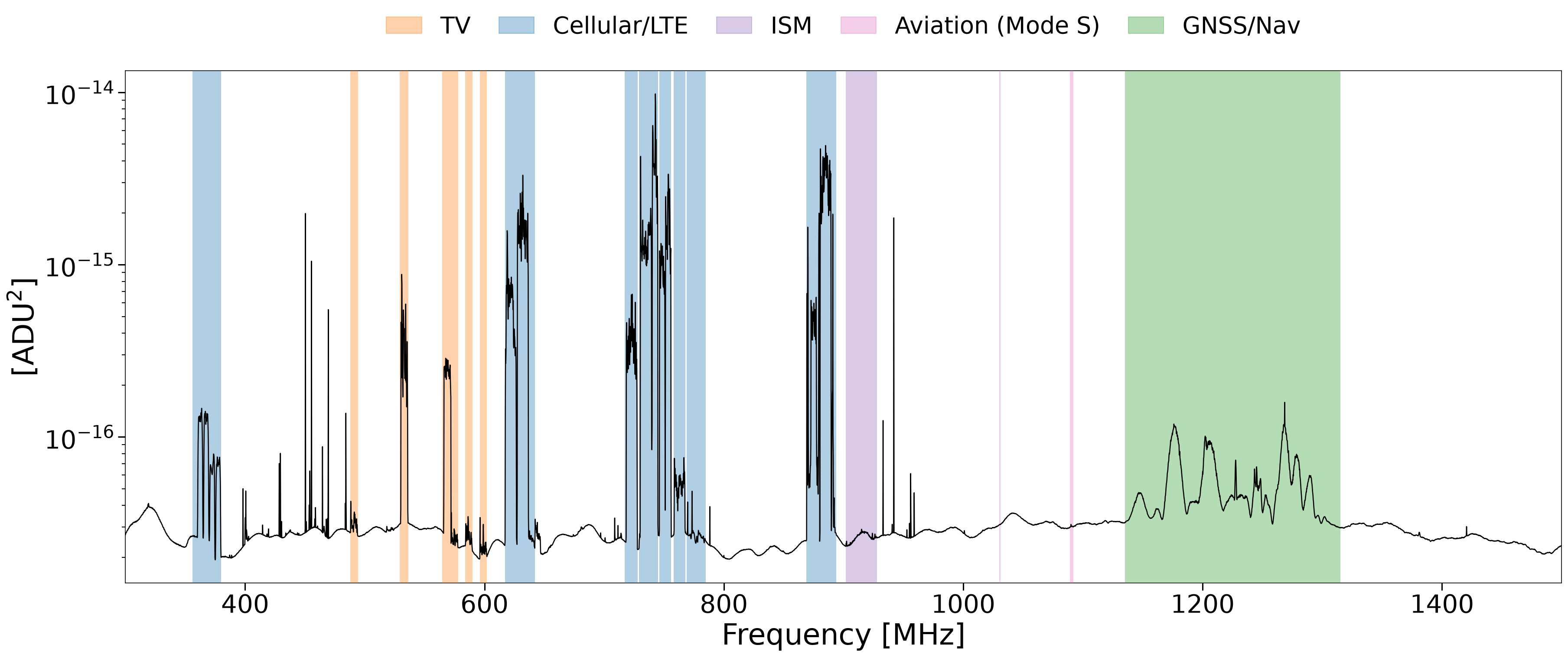}
    \caption{Averaged CHORD autocorrelation from one example antenna and polarization over 5 hours of integration time, with the most ubiquitous known sources of RFI labeled as colored bands. The most common contaminants are cellular bands, TV stations, and global navigation satellites.}
    \label{fig:placeholder}
\end{figure}

A typical example of the RFI environment at the CHORD site and frequency range is shown in the three panel plot in \autoref{fig:rfispec}. The figure shows 5 hours worth of autocorrelation data from one CHORD dish. In top-down order, the autocorrelation waterfall over 300--1500~MHz, the average spectrum of the waterfall, and the spectrum variance of the waterfall are shown. There are notable bands of persistent RFI, such as the Global Navigation Satellite System (GNSS, approximately 150~MHz contaminated at around 1200~MHz) and cellular bands. The major sources of known RFI across 300--1500~MHz at the DRAO site are labeled in \autoref{fig:placeholder}. The largest occupants are cellular bands, TV stations, and global navigation satellites. However, there are also many sources of transient RFI that do not occupy the full 5 hours. The most prominent structure in the waterfall highlights the bright, persistent RFI, while the spectrum variance highlights the variable nature of some of the  sources. Bands like GNSS that have variation due to satellite visibility also show high variance.

The detection and mitigation of these transient RFI sources is paramount for CHORD to deliver on its science goals. CHORD's primary targets involve high-redshift HI intensity mapping, surveys of HI galaxies, and study of fast radio transients, such as fast radio bursts (FRBs) and pulsars. Due to the spectral line nature of HI intensity mapping experiments, RFI causes loss of redshift information for the HI applications. Each frequency channel in an HI mapping experiment probes a different redshift---therefore, a loss of frequency information due to RFI is a cosmic time that becomes inaccessible at that site. Even if a frequency channel contains low duty cycle, ``transient" RFI, if the time samples occupied cannot be identified accurately, the cosmology measurements will still be impacted. Faint RFI that is difficult to flag has been shown to leak into cosmology datasets, contributing to the systematic floor of the measurements. Many major HI cosmology experiments are facing systematic limitations that include potentially unflagged RFI  (\citealt{Wilensky_2023}, \citealt{Mertens_2025}, \citealt{Abdurashidova_2022}, \citealt{Munshi_2024}). While little can be done about persistent, high duty cycle RFI, flagging transient RFI is necessary for successful HI cosmology measurements. 

For radio transient science, RFI can mimic the signals from FRBs or pulsars in the first stage detectors, causing an excessive number of false positive flags per day in the detector (\citealt{Rafiei_Ravandi_2023}, \citealt{rajwade2024}). These false positives are usually separated from real transients in later analysis stages, leading to an excess burden on those pipelines. Additionally, these false positives result in adjustment of the trigger thresholds for the FRB systems to be much higher than would be ideal; otherwise, they would be swamped by these false positives. High trigger thresholds mean that faint signals are likely being missed in the triggering system, and better discernment between RFI and transients would therefore increase the number of faint sources detected. Every science case for CHORD requires carefully-designed RFI identification techniques. The work discussed here will focus on flagging of ground-based, transient RFI in the context of CHORD cosmology, but the techniques and algorithms can be extended with future work to the FRB/pulsar pipelines as well. The concept and technique is also broadly applicable to other radio observatories facing similar challenges. 

RFI monitors are developed for a wide variety of purposes. Site-specific setups target monitoring and archiving the general RFI environment at a specific observatory. These setups often serve multiple instruments, such as the general-purpose monitor for the DRAO (\citealt{Bruce_2026}), or specifically tuned for one instrument, such as that developed for the Giant Metrewave Radio Telescope (GMRT, \citealt{GMRT}). These examples provide information to the observatory through separate data products that allow a study of the RFI environment independent of the telescope. Other instruments such as the Australian Square Kilometre Array Pathfinder (ASKAP) conduct reviews of the RFI environment in the telescope data itself (\citealt{asksp}). Generalized monitors are also often developed for assistance with site selection, such as that designed for the Deep Synoptic Array (\citealt{dsasite}). While these instruments provide invaluable information to the observatory, they are not designed to directly correlate with the science instruments themselves, as they typically do not share a synchronization clock or data pipeline. The intent for the RFI monitor described here is to provide correlated input to the CHORD instrument, directly offering auxiliary flagging information that allows lower levels of contaminated data to be flagged at the same cadence as science data is recorded.

Auxiliary antenna-assisted RFI mitigation is not a new concept, but much of the previous literature focuses on tracking and mitigating specific sources of interference, as well as adaptive cancellation, rather than their use as broad coverage supplementary RFI flaggers for cosmology instruments. \citet{auxant}, \citet{Mitchell_2005}, and \citet{Barnbaum_1998} all focus on the development of auxiliary antennas for adaptive cancellation of a specific interferer. By using a high-gain reference antenna independent of the telescope to measure a specific RFI source, such as FM radio or a satellite, the resulting voltage timestream can be corrected for the signal in real-time, allowing the telescope to ``see through" the RFI. Here, a similar concept is presented, using auxiliary antennas directed at the horizon to provide a high-gain input to the telescope. However, rather than focusing on a specific source, the monitor is intended to provide input across the telescope's entire frequency range and offer horizon spatial coverage. The data streams from the RFI antennas are directly correlated in real time with the CHORD dishes, providing a synchronized measure of the RFI environment.

\begin{figure}
    \centering
    \includegraphics[width=.9\linewidth]{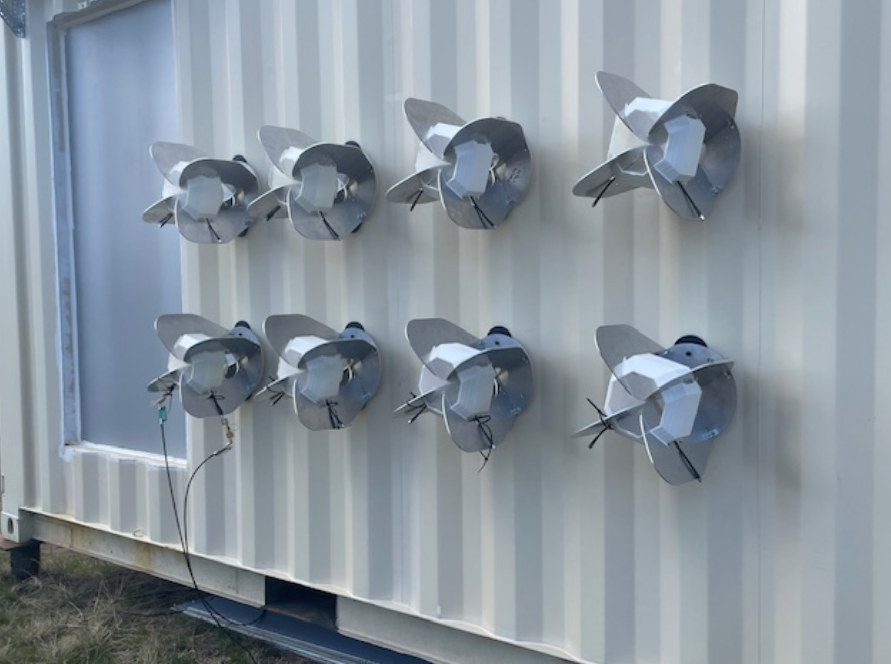}
    \caption{The RFI monitor prototype located at the Dominion Radio Astrophysical Observatory (DRAO). The instrument uses the same feeds as the CHORD instrument. It is mounted on the East side on one of the CHORD receiver huts, arranged into a 4 by 2 array with 56-cm spacing between feeds. The photo was taken before all the feeds were outfitted with electronics for visual clarity. A signal amplification and filtering chain can be seen installed on the lower-left feed.}
    \label{fig:feedphoto}
\end{figure}

This work will describe the design and initial results of an auxiliary array of RFI antennas that cross-correlate in real-time with the CHORD pathfinder dishes. Preliminary data from a single RFI antenna version cross-correlated with the CHORD precursor, the Deep Dish Development Array (D3A) are also shown. Results produced from two different algorithms are shown, each operating over a different timescale. The short-timescale case (unintegrated frames) is produced by a spectral kurtosis estimator (\citealt{Taylor_2019}) and uses the D3A data, and the long timescale case (flagging full 10-second integrations) is produced with the CHORD dish data and a median absolute deviation (MAD) flagger. Both cases show markedly increased flags produced with the inclusion of the RFI monitor, in contrast to flags produced with no additional information from the RFI monitor.  

\section{RFI Monitor Design}

While this work is not focused on evaluating specific RFI flagging algorithms, a brief discussion of them is merited due to the use of multiple different algorithms in evaluating the performance of the RFI monitor. RFI detection algorithms generally fall into two broad categories---online RFI identification, which often happens at short time cadences close to the native time resolution of the telescope (e.g., \citealt{buch2025}, \citealt{van_Nieuwpoort_2016}, \citealt{Dumez_Viou_2016}), and offline RFI identification (e.g., \citealt{Offringa2012}, \citealt{Wilensky_2019}, \citealt{offringa2010lofarrfidetectionpipeline}), which often happens on the final integrated data products, whose time resolution is typically much longer than the sampling time resolution of the telescope. Both timescales typically only use information from the telescope feeds that are also used for science. Many observatories use a combination of these two techniques. The difference is largely due to the limiting factor on how much raw data can be saved to disk. Large N radio telescopes like CHORD cannot afford to save data to disk at the raw time resolution-- they must integrate the data before writing to disk. The benefit of flagging in real time is the capability to do fine-timescale excision before this stage occurs, since RFI often occupies only a small fraction of the time-frequency samples that go into an integration. If this RFI can be excised at time scales much shorter than the integration time written to disk, it is possible to salvage frequencies with low-duty cycle RFI contamination. CHORD will follow in the footsteps of CHIME with the implementation of a real-time, short-timescale spectral kurtosis-based flagger (\citealt{Taylor_2019}) that excises data before the longer integration stage. The expected timescale are 1ms and 40ms excision stages, compared to the expected cosmology integration times of many seconds. However, offline RFI flaggers will always be necessary to identify the RFI which remains after this step. These offline flaggers can often run more complex algorithms as they are not required to do so in real time, and low level, high duty cycle RFI that integrates up over the file writing timescale may be possible to find at the integrated cadence. 

Regardless of which algorithm or timescale is chosen, any RFI flagging algorithm will detect high-significance events more reliably than RFI that lies close to the noise floor of the telescope. This presents a conundrum---radio telescopes are designed to minimize RFI leakage into the instrument, but faint RFI events are more difficult to detect once they are in the data. For a dish-based telescope like CHORD, the problem is even more severe. Much of the RFI will come from the horizon, but the dish attenuates this from view (see \citealt{MacKay_2023} for details on the CHORD beam), resulting in extra attenuation of the RFI. The instrument that will be described in this paper serves to augment CHORD's horizon view with auxiliary antennas, providing a high-gain measurement of the horizon RFI. While many such monitors have been designed for observatories, the novelty in our approach lies in its integration  into the CHORD backend, enabling real-time correlation between the RFI monitor and CHORD antennas. This correlation is key to the goal of being able to flag faint RFI events.

The primary driving specification on the RFI monitor is higher gain at the horizon than CHORD. For the prototype, we targeted a higher gain than the CHORD 64 dish pathfinder instrument. The primary horizon-blocking aspect of the CHORD antennas comes from the dish itself, which results in a beam that only has sensitivity towards the horizon in the sidelobes. The feed itself is fairly omnidirectional, at only 3--7~dBi across the band (see Appendix A). Therefore, the bare CHORD feed was chosen as a test case, as one single CHORD feed pointed towards the horizon should outperform any one single CHORD dish in the horizon direction. The approximate gain from beamforming $N_\text{ants}$ antennas is

\begin{equation}
    G = 10*\log_{10}(N_{\text{ants}}).
\end{equation}

\begin{figure}
    \centering
    \includegraphics[width=.7\linewidth]{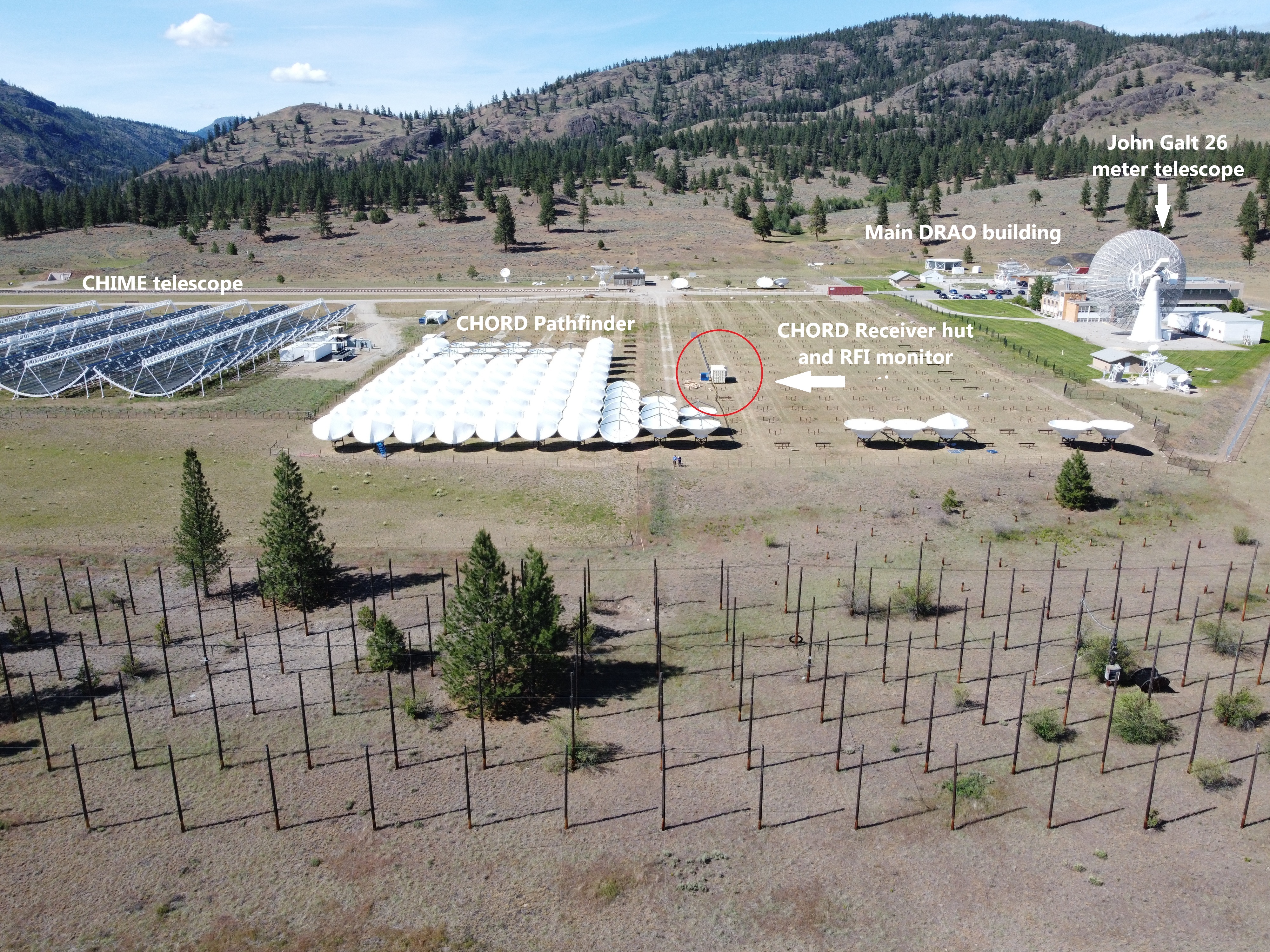}
    \caption{Location of the receiver hut where the RFI monitor antennas are mounted. The antennas are looking east, away from the CHORD pathfinder and towards the Galt 26-meter telescope and main DRAO building. Image supplied by DRAO, HAA, National Research Council of Canada.}
    \label{fig:layout-a}
\end{figure}

\begin{figure}
    \centering
    \includegraphics[width=.7\linewidth]{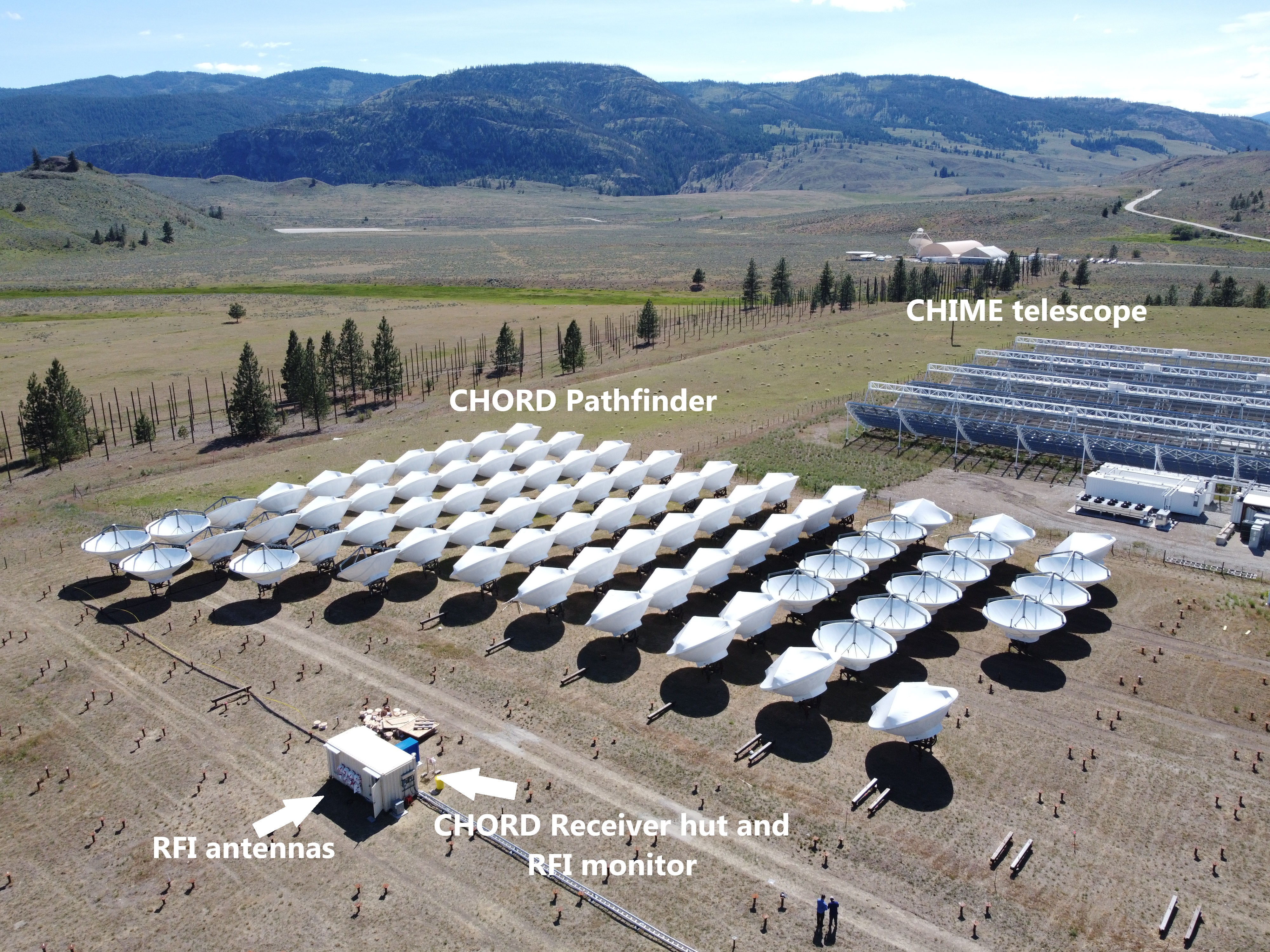}
    \caption{A view of the receiver hut with the RFI antennas visible.  Image supplied by DRAO, HAA, National Research Council of Canada.}
    \label{fig:layout-b}
\end{figure}

\noindent

The sensitivity of a single CHORD antenna is around $-$20 to $-$10~dBi at the horizon, depending on the frequency, with combining signals from the 64 dish pathfinder array adding 18~dB ($-2$ to 8 dBi at the horizon), and the full 512-dish instrument adding 27~dB (7 to 17 dBi at the horizon). In order to surpass this we need one feed for the single-antenna comparison, 8 for the pathfinder, and 16 for the full CHORD core array. 

For this prototype, we chose to construct the 8-feed version shown in \autoref{fig:feedphoto}. While the majority of the analysis in the paper will be done with 1 feed and 1 dish, the 8 antenna version allows for future investigations once the full CHORD pathfinder is online. The antennas are located on the side of the ``receiver hut'' which sits in the array footprint. The hut is a customized steel shipping container hosting the second-stage amplifiers, digitizer boards, and GPU nodes for the CHORD pathfinder. The array is facing east towards the main buildings at the DRAO. \autoref{fig:layout-a} and \autoref{fig:layout-b} show the location of the container. The feeds are attached to the side of the shipping container with off-the-shelf magnets, and spaced 56~cm apart, largely set by the corrugations on the container and physical size of the feeds. The 4 by 2 layout serves to allow future investigations of RFI directionality. Currently, all inputs are digitized separately, but the vertical direction could be reduced in the future with an analog sum to reduce the number of digitizer inputs needed. Due to the fact that the current iteration of the monitor has significant ground pickup in the beam, narrowing of the beam in the vertical direction is desirable for noise purposes as well as reduction of inputs, and future versions will likely include vertical analog beamforming.

\begin{figure}
    \centering
    \includegraphics[width=.8\linewidth]{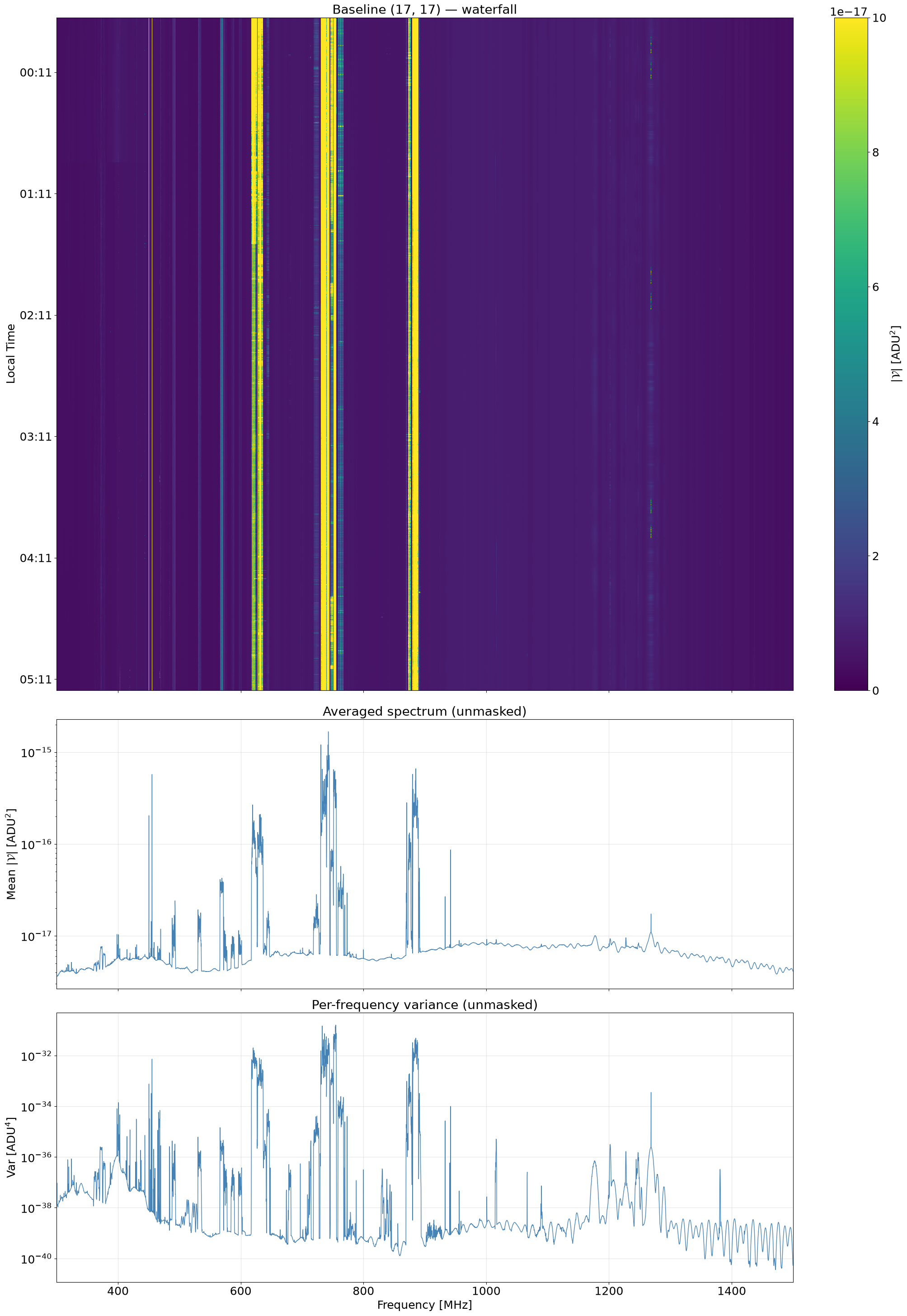}
    \caption{Waterfall, averaged autocorrelation spectrum, and spectrum variance for one polarization of one antenna of the RFI monitor. The response of the RFI system varies from the CHORD system due to the lack of a dish, different low noise amplifiers, the uncorrelated polarization, and the fact that the feeds are seeing the horizon instead of the sky.}
    \label{fig:rfimonex}
\end{figure}

While some RFI is expected to have polarized emission (e.g, \citealt{Li_2023}), the polarization is difficult to predict due to the variety of possible sources as well as reflection and geographical effects. The polarization of the antennas was set by the configuration that allowed stability on the container and did not require any extra holes to be drilled in the feed backplane for mounting, and therefore is not expected to be aligned with the polarization of the CHORD dish feeds. The feeds have a 30$^\circ$ offset from perpendicular. This paper will refer to the slightly more horizontal polarization as X, and the slightly more vertical one as Y. The estimated error bar on the baseline lengths and polarization angle is $\pm$1~cm in baseline and $\pm 1 ^{\circ}$ in angle. The limitation was due to human error on positioning the magnets.

The analog signal chain uses many of the CHORD components (see \citealt{chord2026}). The low noise amplifier (LNA) is an off-the-shelf Mini-Circuits component (ZX60-P103LN+, \citealt{minicircuits_zx60p103ln}), and the first- and second-stage filter linear amplifiers (FLAs) from the CHORD design are used for the RFI monitor signal chains. The signals are digitized separately on 16 inputs (8 RFI antennas, dual-polarization) on two t0.technology Control and Readout System (CRS) boards (\citealt{Montgomery2024CRS}, \citealt{HendricksenCHORDFEngine2026}). These boards connect and sync directly to the CHORD backend, enabling real-time cross-correlation with the array. 

An example autocorrelation waterfall from one polarization of one RFI monitor antenna, as well as the averaged spectrum and variance are shown in \autoref{fig:rfimonex}. Differences in the spectrum from the CHORD dish autocorrelations are expected due to the bare feed setup, the different LNA used, the lack of correlation in polarization between the two, and the fact that the feeds are pointed at the horizon and not the sky.

\section{Results}

\subsection{Fine-Timescale Spectral Kurtosis Flagging} \label{sec:fts}

\begin{figure}
    \centering
    \includegraphics[width=\linewidth]{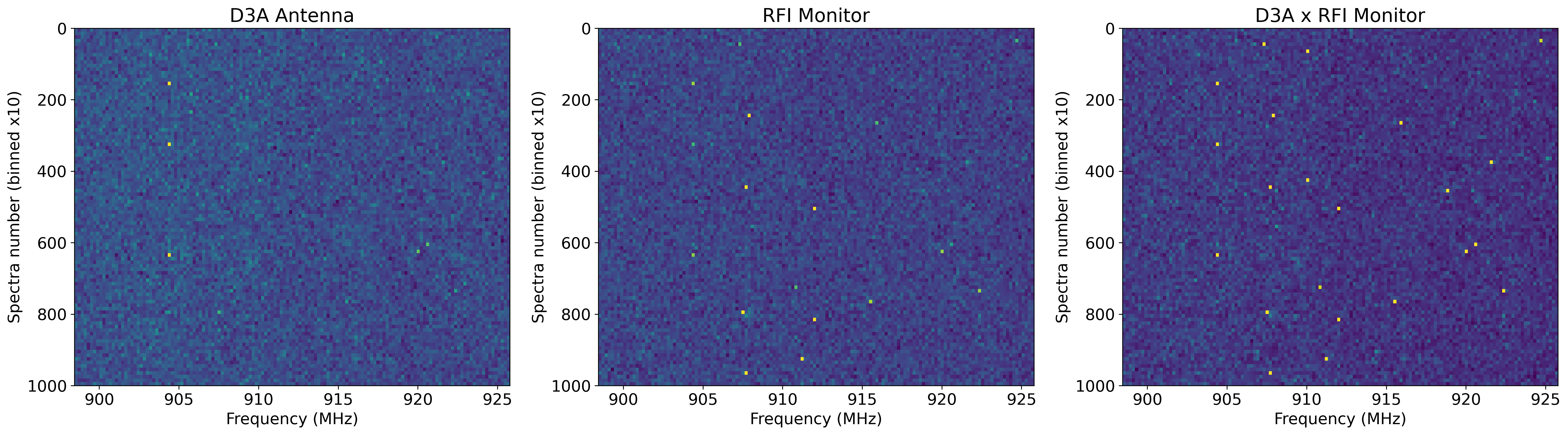}
    \caption{Single D3A dish and single RFI monitor feed waterfalls along with the cross correlation between the two, zoomed in to 1000 spectra and a 30-MHz range covering a 902--928~MHz ISM band, for the data used to produce the subpanel flags shown in \autoref{fig:flags}. The data is binned by a factor of 10 for visual clarity. (Left panel) D3A antenna waterfall (Middle panel) RFI monitor waterfall. (Right panel)  Cross correlation between the RFI monitor and D3A antennas. The third panel is shown as the RFI is more obvious by eye in the cross correlation between the two data sets, but the spectral kurtosis estimator used runs over raw voltage streams so this product is not used to produce flags. The short duration, bright RFI in the cross correlation offers an ideal target for short timescale RFI excision. Eliminating bursts of data around the bright pings could salvage this otherwise unusable ISM band for science use.}
    \label{fig:ftscomp}
\end{figure}

\begin{figure}
    \centering
    \includegraphics[width=\linewidth]{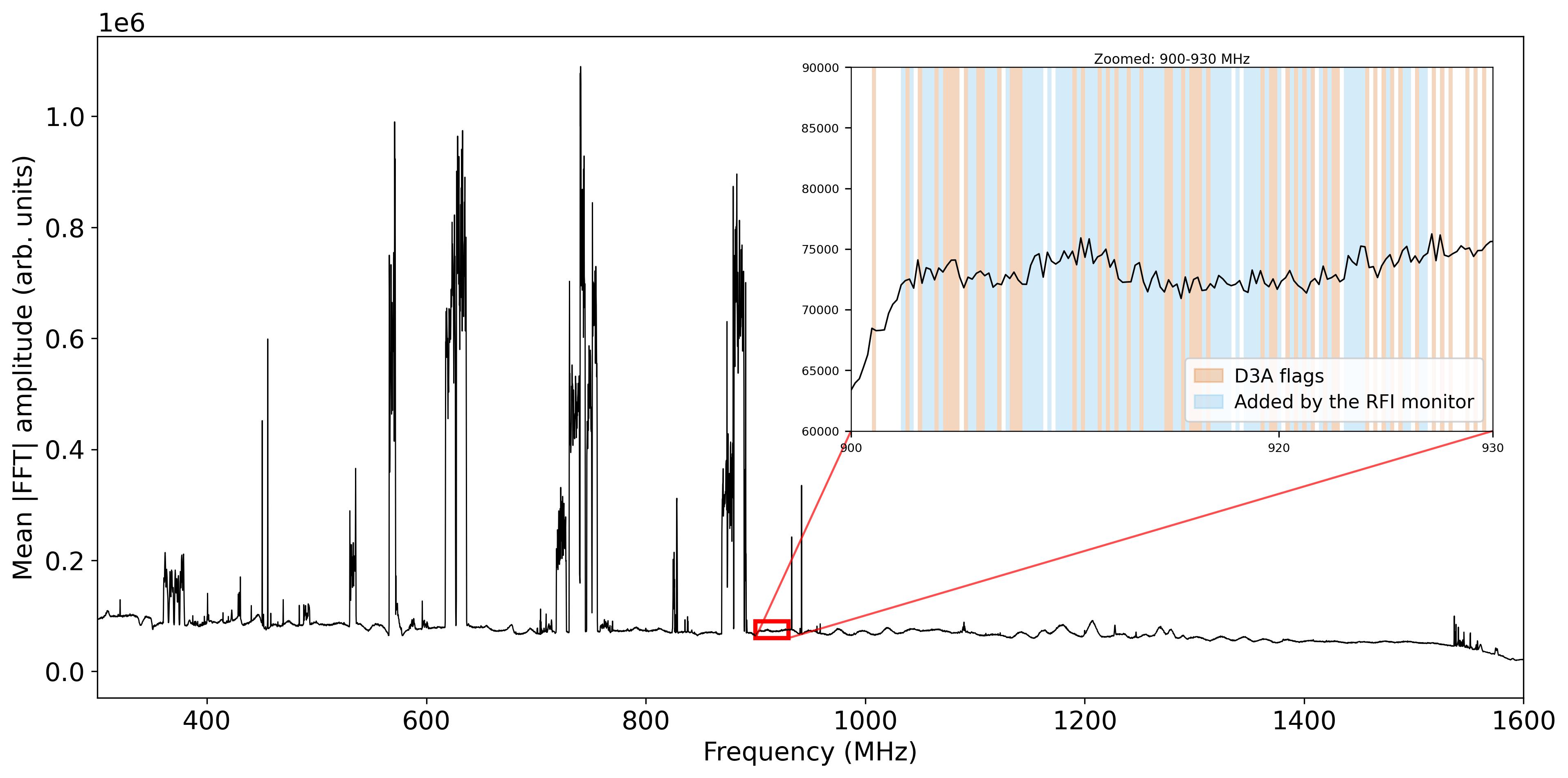}
    \caption{Averaged spectrum over 5000 captures, with an inset panel zoomed in on 900-930 MHz, covering a 902--928~MHz ISM band. The zoomed panel shows the flags produced with spectral kurtosis for a 5000-spectra data capture, with the flags produced directly from the D3A data in orange and the blue indicating additional flags added by the RFI monitor. D3A alone flags 51/154 channels (33.12\%) and the RFI monitor's flags are a superset, with 44.16\% of the band only flagged by the RFI monitor. The remaining 22.73\% (35 channels) are unflagged.}
    \label{fig:flags}
\end{figure}

Flagging samples using RFI antennas sampled synchronously at the native data rate of the telescope offers a powerful way to preserve longer integrations from contamination and prevent data loss. These short-timescale flaggers directly remove samples before they are integrated into a longer spectrum. The native data rate of each ADC in the CHORD telescope is 3.2~GSPS per antenna, with individual Fourier-transformed frames produced every 5.12~microseconds. The total frequency coverage of the digitized data is up to 1600 MHz with a spectral resolution of 195.3 kHz, but this is downselected to 300-1500 MHz before being transported off the board. For low-duty cycle RFI, removing just a small amount of the data before integrating can prevent entire integrations from needing to be flagged later. The RFI monitor described here can provide a secondary input to these real-time flaggers with higher sensitivity towards sources at the horizon, potentially detecting RFI that is below the threshold for a particular flagging method in the main telescope, but still impactful enough to contaminate the science.  Spectral kurtosis is a popular short-timescale flagging algorithm---the natural astronomical processes can be modeled as a zero-mean Gaussian distribution when they evolve slowly compared to the window over which SK is computed, and looking at the higher moments of the field offers a powerful way to identify non-Gaussian RFI sources. The spectral kurtosis timescales for CHORD are expected to nominally be computed over 1~millisecond to 40~millisecond timescales, and the total integration time written to disk will be 10--30 seconds of data. 

The initial test of fine-timescale flagging was done with a precursor to the CHORD instrument, called the Deep Dish Development Array (D3A, \citealt{kit_msc_thesis, Hendricksen2023, aditya_msc_thesis, vincent_phd_thesis, elizabeth_msc_thesis}), and a precursor to the RFI instrument that was a single CHORD feed version. At the time of the analysis presented here, baseband data at the native time resolution of the telescope was unavailable from the full CHORD array, so a reduced dataset was taken with the eight inputs that could fit on a single digitizer board, on May 12th, 2025. The $1$~GbE network bandwidth on the CRS board in this testing configuration also limited the throughput of the frames in this reduced configuration. 2 frames were captured every 10 milliseconds, for a total of 5000 frames. Therefore, the fine timescale RFI shown in the following figures has significant gaps. However, we expect the sky signal statistics to remain stable over this time period, and deviations from Gaussianity due to RFI to still be present. This analysis was done offline, but represents a similar analysis to what could be done in real time in future analyses.

For a realistic example of fine-timescale flagging, here we implement a spectral kurtosis-based flagger similar to that implemented for CHIME (\citealt{Taylor_2019}).
The spectral kurtosis is computed by taking the per-frequency second and fourth moments over n time samples. For each feed i:
\begin{equation}
\mu_{2,i} = \frac{1}{n}\sum_{t=1}^{n} |\tilde{E}_{ti}|^2
\qquad
\mu_{4,i} = \frac{1}{n}\sum_{t=1}^{n} |\tilde{E}_{ti}|^4
\end{equation} 

and calculating:
% Spectral kurtosis for feed i:
\begin{equation}
\text{SK}_i = \frac{n+1}{n-1}\left[\frac{\mu_{4,i}}{\mu_{2,i}^2} - 1\right]
\end{equation}

For a complex Gaussian distribution we expect 

\begin{equation}
\langle \text{SK} \rangle = 1
\qquad
\text{Var}(\text{SK}) \equiv \sigma^2 = \frac{4}{n}
\end{equation}

We flag any 5$\sigma$ outliers from the expected Gaussian distribution. The flagging timescale was 5000 spectra, comprising 25.6 milliseconds worth of integrations. However, due to the network constraints, these captures occur over 25 seconds of actual time. The expected integration times for the actual CHORD array are 10--30 seconds, so this is roughly representative of the timescale of data that would go into one integrated spectrum. \autoref{fig:ftscomp} shows waterfalls from this dataset with a reduced frequency  and time subset to highlight an ISM band occupying 902-928 MHz. The spectra are binned by a factor of 10 for visual clarity, but the spectral kurtosis algorithm was run on the raw data. The left waterfall shows the spectra from the D3A antenna, the middle panel from the RFI monitor, and the rightmost panel the cross-correlation between the two. Bright, speckly pings are obvious in all three waterfalls, but have much higher signal to noise in the RFI monitor and the cross-correlation spectrum. 

The spectral kurtosis flags produced from the 5000-sample set (a subset of which was included in the waterfalls in \autoref{fig:ftscomp}) are shown in \autoref{fig:flags}, highlighting the portion of the band flagged with the D3A data and the RFI monitor. The flags are shown in the zoomed panel over 900--930 MHz. The flags produced from the D3A data alone occupy 51/154 of the channels (33.12~\%). The flags produced with the RFI monitor are a superset of those produced with D3A---the flags occupy the same 51 channels and an additional 68 channels. 44.16~\% of this range is occupied by flags only produced by the RFI monitor. 22.73~\% of this range (35 channels) remains unflagged. The cross-correlation flags are not shown, as the spectral kurtosis estimator used here is computed over raw voltage streams rather than cross-correlations. Cross-correlation spectral kurtosis estimators have been developed (\citealt{2020ursi.conf..235N}) and should be investigated for other frequency bands where the extra boost in sensitivity would be beneficial. The very short burst nature of the RFI detected with the spectral kurtosis algorithm also opens the possibility of investigating the excision of even shorter bursts than the 5000-frame example here, salvaging this previously unusable band for science.

\begin{sidewaysfigure}
    \centering
    \includegraphics[width=.9\linewidth]{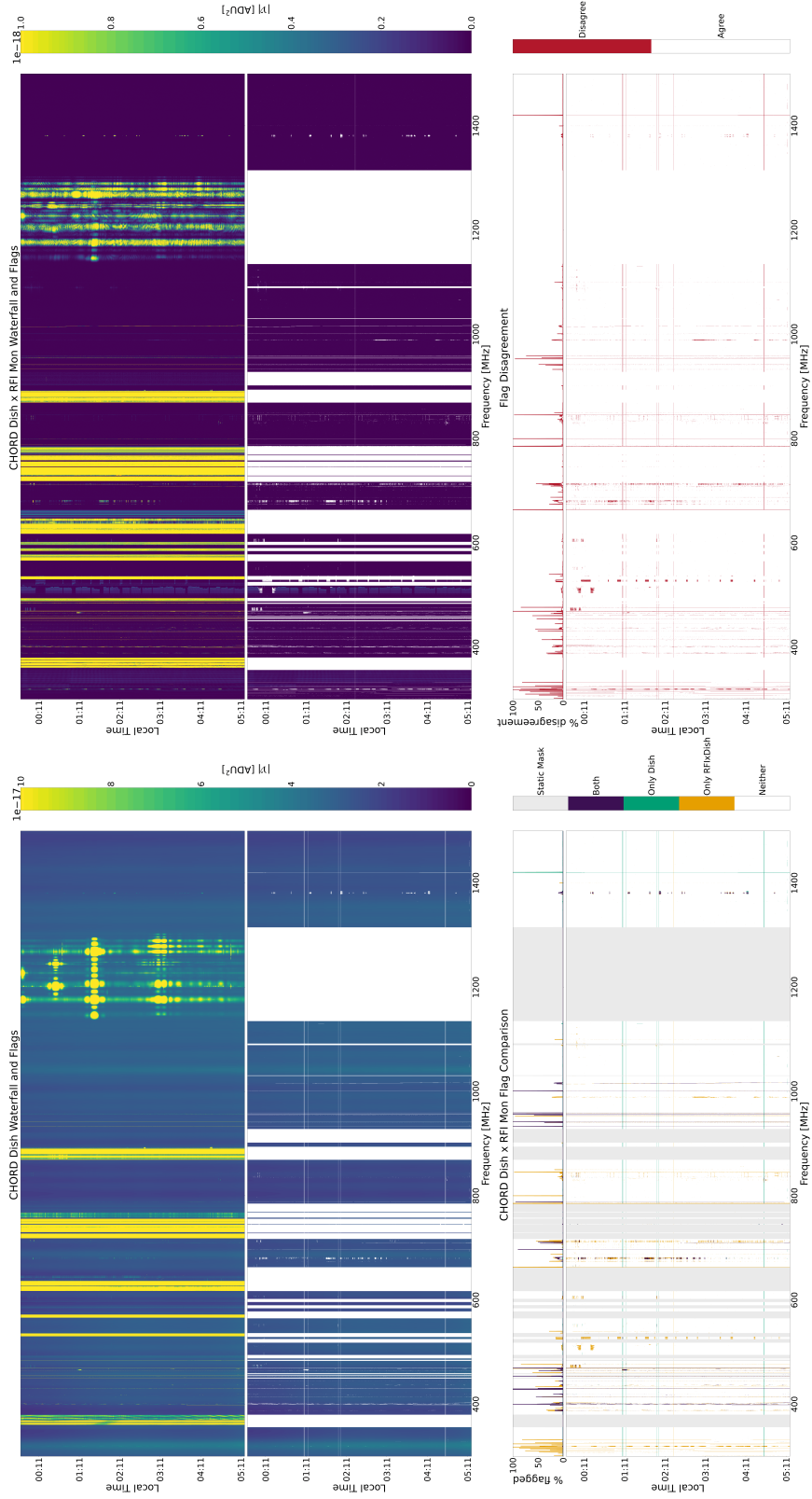}
    \caption{Top Left: CHORD Dish autocorrelation waterfall and set of MAD flags, with 35.858\% of the data masked. A static mask is applied that covers 35.091\% of the data, to address channels with persistent, bright RFI. Top right: CHORD dish and RFI monitor cross-correlation and set of MAD flags. The static mask covers 35.091\% of the data. 36.389\% of the data is masked after the MAD flagger. Bottom Left: Flags colored by product, with flags that occur in both the top plots in purple, flags that show up for only the CHORD dish in teal, and flags from only the RFI monitor and dish cross-correlation in orange. The static mask is plotted in grey. The RFI monitor and dish cross-correlation produces a significantly increased flagging fraction, with 26.45~\% of frequencies having at least one time sample that is masked only in the cross-correlation. Bottom Right: Difference plot between the CHORD dish flags and the RFI monitor and CHORD dish cross-correlation.}
    \label{fig:fullpage}
\end{sidewaysfigure}

\subsection{Integrated correlation flagging} \label{sec:intcorr}

\subsubsection{Single Antenna Case}
While fine-timescale flagging offers a powerful way to save more integrated data, spectral kurtosis is not expected to catch 100\% of RFI. CHORD cosmology will also use offline flaggers at the file-writing cadence, expected to be 10--30 seconds, to catch RFI that is not excised by the shorter timescale stages. This helps prevent contaminated integrations from being included in the cosmology analysis. Traditional packages for doing this include algorithms like the Sky Subtracted Incoherent Noise Spectrum (SSINS, \citealt{Wilensky_2019}) and AOflagger, based on a sumthreshold method (\citealt{Offringa_2010}). Here, we test with a Median Absolute Deviation (MAD) flagger applied after a median subtraction to remove the background sky.

The prototype signals were digitized and channelized with the same digitizer boards as the CHORD dishes and directly plugged into the CHORD pathfinder backend, enabling real time cross-correlation with the pathfinder CHORD dishes over the full 4 by 2 RFI array. Using the cross-correlations is powerful, as the lack of overlap in field of view between the CHORD dish and the RFI antennas means the primary source in common is the shared RFI environment-- the cross-correlation strongly suppresses receiver noise and the sky. The observation totals approximately 5 hours worth of data taken on April 24th, 2026. At the time of the observation, 3 dual-polarization CHORD dishes were in operation, and the dataset spans roughly 5 hours of time during the night at the DRAO. One polarization of one dish and one polarization of one RFI monitor feed were chosen as test cases. A static mask was manually applied over approximately 35\% of the data, covering known persistent RFI sources. A MAD flagger was run on the dish autocorrelation, as well as the dish and RFI monitor cross-correlation. The flagger ran as a sliding 2D window over 3 iterations, identifying 5 sigma outliers at each stage. Each integration is 10 seconds of data, and the MAD flagger produces per-integration, per-frequency flags. 

A summary of the flagging statistics is shown in \autoref{fig:fullpage}. Top-left shows the dish autocorrelation and flags, top-right shows the dish cross-correlation with the RFI monitor, bottom-left shows the flags compared for each product, and bottom-right shows the difference of the flags. Numerically, the static mask occupied 35.091\% of the data, the mask on the dish autocorrelation was 35.858\% (+0.767\% from the static mask), and the RFI monitor and dish cross-correlation was 36.389\% (+1.298\% more than the static mask). The RFI and dish cross-correlation is thus flagging 70\% more non-static masked data than the CHORD dish alone. Although numerically the total frequency- and time-occupancy of the extra flags is low, at the percent level, 26.45\% of frequency bins show the CHORD dish and RFI monitor cross-correlation flagging at least one time sample that is unflagged by the dish data alone. Even a small amount of missed RFI can contaminate an entire frequency bin for cosmology analysis, and this elimination of the extra 1.298~\% of samples represents a large impact across the CHORD frequency range.

The difference plot shows a few features that are only flagged in the dish autocorrelation. This is not an unexpected result---the dishes are highly directional, and while we expect much of the RFI to come from the horizon, sources that are in the field of view of the telescope, such as airplanes or satellites, will naturally appear strongly in the dish autocorrelation. Additionally, the current implementation of the RFI monitor is directed east, with a large steel container acting as a ground plane behind it. This effectively blocks much of the horizon from the RFI monitor, meaning anything strongly directional to the west is unlikely to be caught. Upon inspection, the few broadband features that are caught in the dish autocorrelation and not the RFI monitor appear to be real broadband RFI of unknown source. 

\begin{figure}[hbtp]
    \centering
    \begin{subfigure}[b]{.8\textwidth}
        \centering
        \includegraphics[width=.8\textwidth]{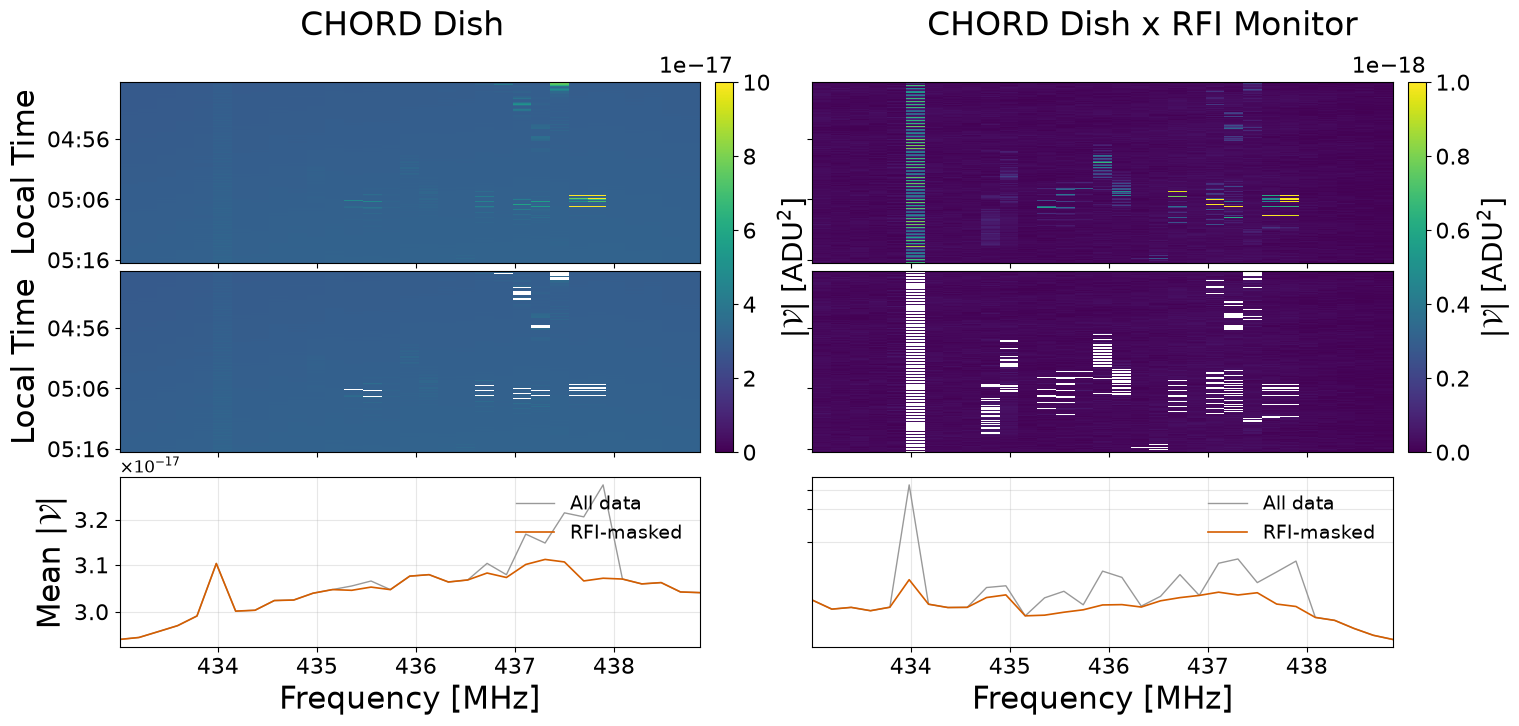}
        \caption{A zoomed in view of the flagging occupancy for 30 minutes of data from 433-439 MHz. A repeating tone at 434 MHz, barely visible in the autocorrelation, is flagged in the cross correlation. Additionally, from 434-438 MHz there is a more broadband event that appears more extended in frequency and duration in the cross correlation.}
        \label{fig:sub1}
    \end{subfigure}
    \hfill
    \begin{subfigure}[b]{.8\textwidth}
        \centering
        \includegraphics[width=.8\textwidth]{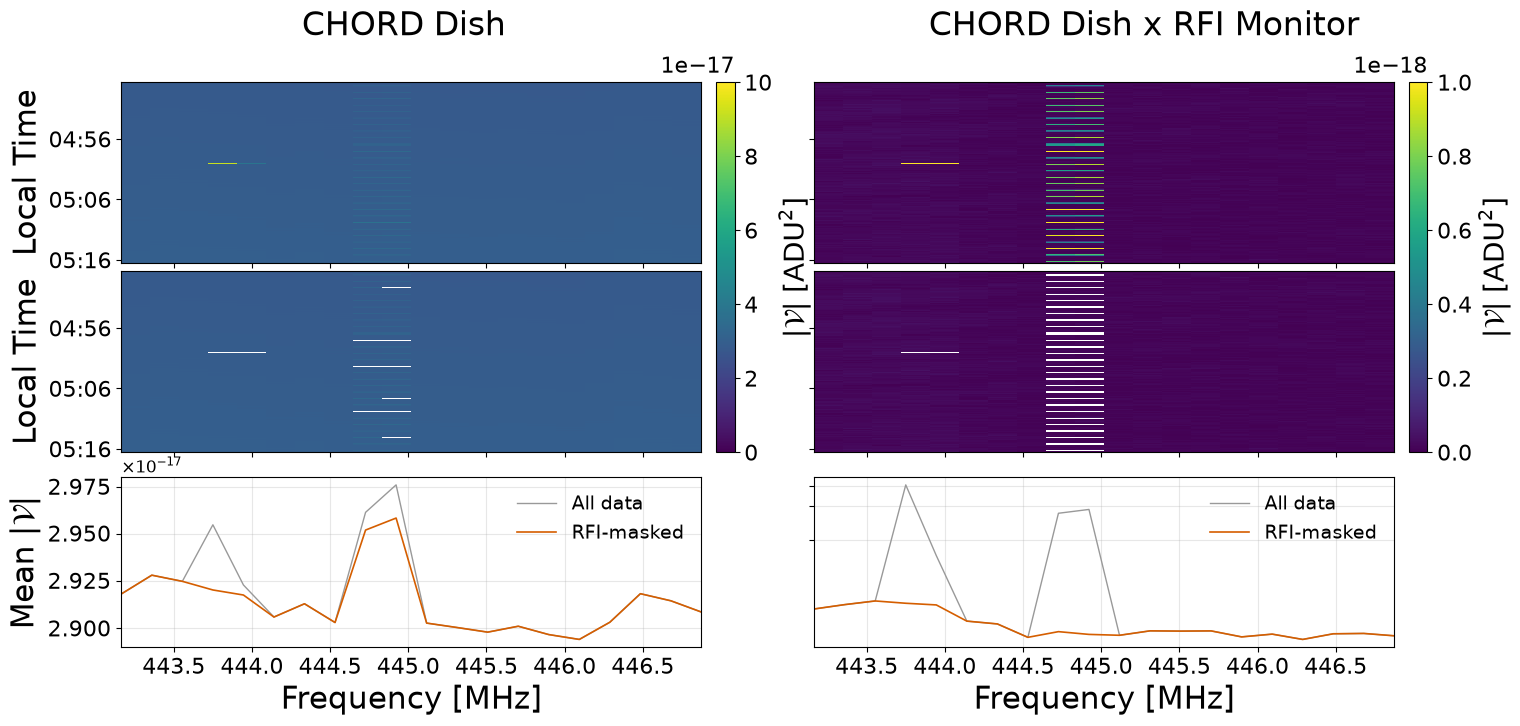}
        \caption{A zoomed in view of the flagging occupancy for 30 minutes of data from 443-447 MHz. A single frame event around 444 MHz is caught by the auto and cross correlations, while the repeating tone around 445 MHz is captured much more clearly in the cross correlation.}
        \label{fig:sub2}
    \end{subfigure}
    \hfill
    \begin{subfigure}[b]{.8\textwidth}
        \centering
        \includegraphics[width=.8\textwidth]{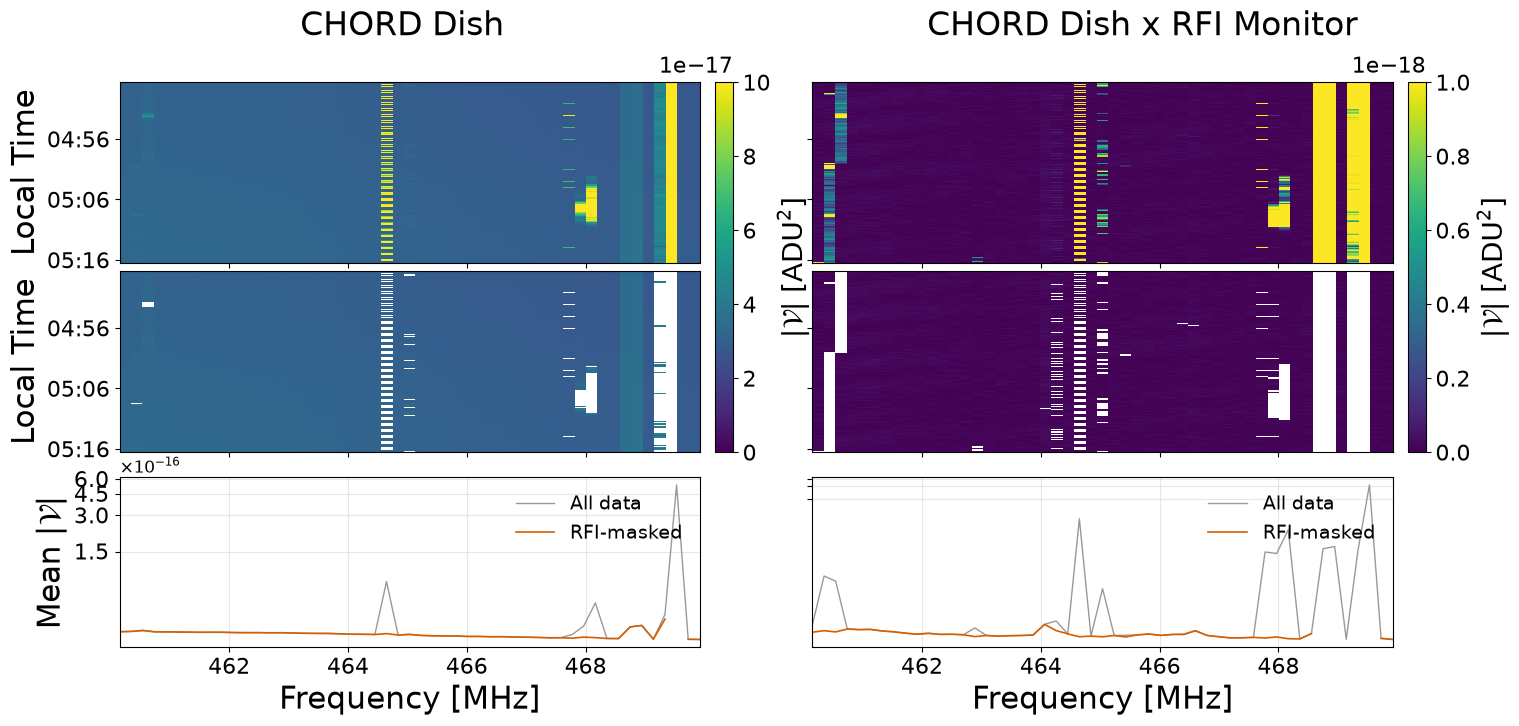}
        \caption{A zoomed in view of the flagging occupancy for 30 minutes of data from 460-470 MHz. The repeating tone around 465 MHz is captured well in both the autocorrelation and the cross, but other features such as the repeating tone just to the right of it, the tone around 469 MHz, and the two at the lower end of the frequency range are only flagged in the cross.}
        \label{fig:sub3}
    \end{subfigure}
    \caption{Zoomed in panels showing a comparison of the flags produced by the CHORD dish autocorrelation and the dish and RFI monitor cross. Many features are captured with the RFI monitor addition that are not apparent or only dimly so in the autocorrelation.}
    \label{fig:zooms}
\end{figure}

\FloatBarrier
The vertical narrowband flags at approximately 1420~MHz in the CHORD dish data correspond to the 21cm line from the Milky Way, and are not present the RFI monitor flag set due to the low overlap in field of view between the two instruments. Without the RFI monitor, it is difficult for algorithms to distinguish bright signals like this one from RFI. By cross correlating between the RFI antenna and the CHORD dish, it is much harder for sky signals to mimic RFI and produce false positive flags.

In this paper, we refer to the flags produced by the RFI monitor as extra flags rather than additional RFI caught. We have no ground truth measure of horizon RFI to compare our flags against. To check that the extra flags can reasonably be attributed to RFI, we inspected them in detail and include a few snapshots here. \autoref{fig:zooms} shows a few close-ups of small time and frequency chunks. In each case, the RFI monitor adds structure that does not look like random inclusion-- it typically extends an event already visible in the autocorrelation either in time or frequency. Figures~\ref{fig:zooms}a and \ref{fig:zooms}b each contain a repeating tone, near 434 and 445~MHz respectively, that is faint and unflagged, or only partially flagged, in the autocorrelation alone, but stands out clearly in the cross-correlations. Figures~\ref{fig:zooms}a and \ref{fig:zooms}c also show broader structures that are obvious in the autocorrelation but appear considerably more extensive in the cross. In these cases, and across the band more generally, the extra flags follow real events plausibly attributable to RFI rather than adding flags at random.

There are possible exceptions to this attribution. The results in this paper focus on CHORD cosmology, where the signals of interest evolve on slow timescales and do not produce sidelobe signals easily mistaken for RFI. Other science cases, however, must account for astrophysical signals entering through the telescope sidelobes that may mimic RFI. For example, CHIME/FRB and STARE2 detected an extremely bright radio burst from the Galactic magnetar SGR~1935+2154 that arrived through the far sidelobes of CHIME (\citealt{2020Natur.587...54C}, \citealt{2020Natur.587...59B}). An event like this in the direction of the RFI monitor would be bright enough to correlate and would likely be flagged as RFI, since nothing in this method distinguishes the two. A future goal of this work is to lower the FRB detection threshold by reducing RFI false positives, and sidelobe events like this would need careful consideration before the technique is applied to the transient pipelines.

\begin{figure}
    \centering
    \includegraphics[width=0.8\linewidth]{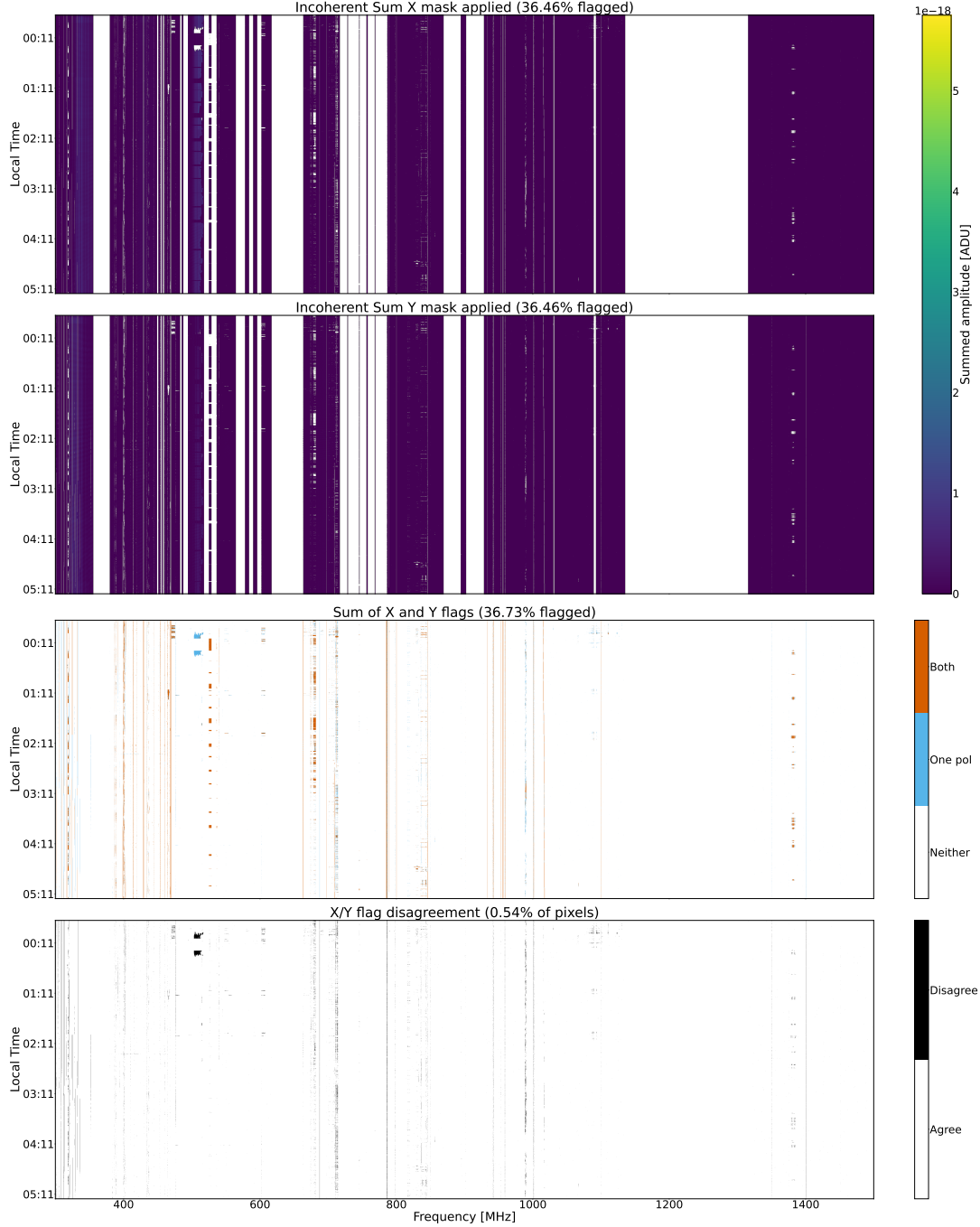}
    \caption{Comparison between the X and Y polarizations of an incoherent sum over all RFI antennas. The top two panels show the flags produced for X and Y respectively, the third panel shows the union of the two flags, and the last panel shows the disagreement for the two polarizations. Both polarizations flag a similar percentage of pixels, but disagree on 0.54\% of those identified as RFI.}
    \label{fig:xycomp}
\end{figure}

\begin{figure}
    \centering
    \includegraphics[width=0.9\linewidth]{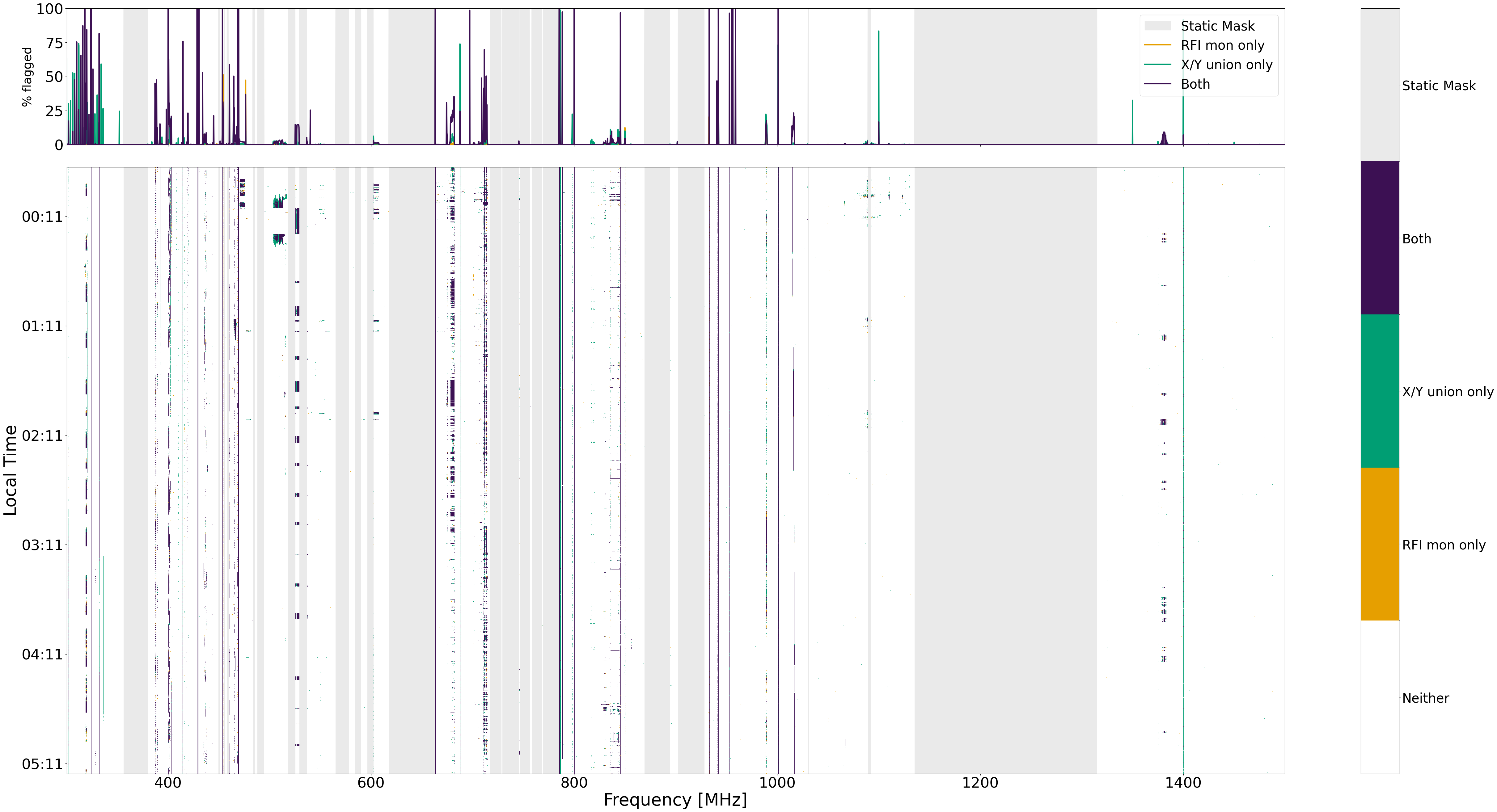}
    \caption{Comparison between the set of flags produced by the single antenna, single polarization RFI monitor case and the union X/Y incoherent sum flags. Most RFI captured in the single antenna case is also captured by the incoherent sum X/Yflags. The union flags add about .44 \% to the single antenna only case.}
    \label{fig:boolplot}
\end{figure}

\subsubsection{Multiple Antenna Case}

In addition to the single-antenna cases, one can use the power of having multiple antennas to increase the sensitivity. Here, all cross-correlations of the 8 RFI antennas with the same CHORD dish used above are incoherently summed for the same polarization, and the X/Y flags are unioned to form a final flag set. This effectively forms a single beam looking directly east, with the same spatial coverage as a single antenna. The signal to noise gain of the incoherent sum is not as high as the coherent beamforming mentioned in section 2-- the SNR improvement now scales by $\sqrt{N}$ rather than $N$. However, when comparing to one single CHORD dish instead of the whole pathfinder the extra beamforming gain is not necessary, and the incoherent sum allows for viewing the whole eastern horizon at once rather than considering a particular direction. Beamformed analyses are left to future work. The polarization of the RFI monitor is not aligned with respect to the polarization of the CHORD antennas, so both the X/Y flags are relevant for any given CHORD antenna and polarization. 
The resulting set of flags is shown in \autoref{fig:xycomp}, with the independent X/Y flags shown over the incoherent sum waterfalls, followed by the union flags and the X/Y flag disagreement. Both polarizations flag a similar percentage of RFI, but disagree on some subset of the flags due to some RFI with polarization dependence. \autoref{fig:boolplot} compares the union incoherent sum X/Y flags to the RFI monitor single antenna flags shown in \autoref{fig:fullpage}. Very little data (0.1\%) is flagged only by the single antenna---this corresponds to one single broadband event that only appears in the one antenna used for the first test case. The X/Y incoherent summed union adds another 0.44\% to the set of flags produced by the single polarization single antenna case. This corresponds to an increase in RFI flags of about 36\%. 20.51\%  of the CHORD frequency range is impacted by adding the incoherent sum X/Y union---there are  1,260 of 6,144 (20.51\%) frequency bins where the X/Y union flags something that the single-polarization, single-input case missed. This represents a strong case for using the full 8-input monitor for the CHORD pathfinder, and scaling up the number of antennas for the full CHORD core array.

\section{Conclusion and Future Work}

\begin{figure}
    \centering
    \includegraphics[width=.8\linewidth]{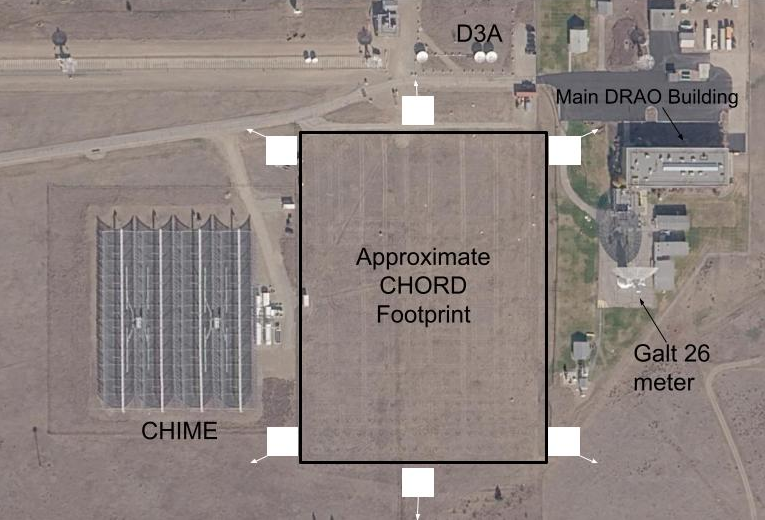}
    \caption{Due to shadowing from the CHORD array and the infeasibility of raising the RFI monitor above the array, therefore putting it into the field of view of the telescope, one option for an upgraded antenna arrangement includes many directional RFI monitors surrounding the array. A potential layout is shown in this figure, indicated by white boxes surrounding the approximate CHORD footprint. The graphics are overlaid on map data from Google Earth.}
    \label{fig:chordint}
\end{figure}

The auxiliary antenna RFI flagging instrument described here provides a powerful way to boost the sensitivity of traditional RFI flaggers to faint RFI originating from the horizon, and make use of the extra information by inputting it to the same correlator as the sky signals. While the design discussed is only a prototype version, it already shows immense promise in supplementing the flags produced for the CHORD instrument. A notable caveat to the results shown here is the current directionality of the RFI monitor. The instrument only sees directly from east because the receiver hut acts as an effective ground plane. In order to obtain true omnidirectionality, a number of monitors would need to be placed around the array. \autoref{fig:chordint} shows a possible layout to increase the number of monitors and provide full horizon coverage. Due to shadowing from the main array, even without a groundplane a true omnidirectional antenna would be largely blind in the array direction, necessitating multiple instruments. One could, of course, elevate an omnidirectional antenna above the main array, but this would require moving the antenna away from the main array to avoid it entering the main beam of the instrument, thereby reducing the likelihood that the two instruments are truly seeing the same RFI environment.

Future work includes a few notable avenues. At the time of the observations in Section~\ref{sec:intcorr}, three CHORD dishes were in operation. The 64-dish CHORD pathfinder is rapidly coming online, with more dishes integrated into the correlator on a weekly cadence. Investigation of the performance of the RFI monitor combined with the full pathfinder will be a critical step in determining the usefulness of this instrument in the full CHORD array. 

Adaptive cancellation, wherein a synthesized beam is constructed to minimize the gain in the direction of measured RFI while maintaining the gain along boresight, presents a compelling future avenue for a system like this. This must be done carefully, as it risks corrupting the underlying science signal due to the effective synthesized beam changing dynamically. The cross-correlation between the RFI antennas and the CHORD dishes also contains the information needed to ``peel" the manmade signals from the voltage streams, allowing the possibility of recovery of the astronomical signal by subtracting the RFI. Investigation into spatially canceling or subtracting RFI using this method is a long term goal of this project. 

Further avenues beyond those noted here may also include those such as investigation into integration with the other CHORD science pipelines, optimization of antenna designs, and development of more general purpose variations to be deployed at observatories other than CHORD.   

\begin{figure}
    \centering
    \includegraphics[width=.8\linewidth]{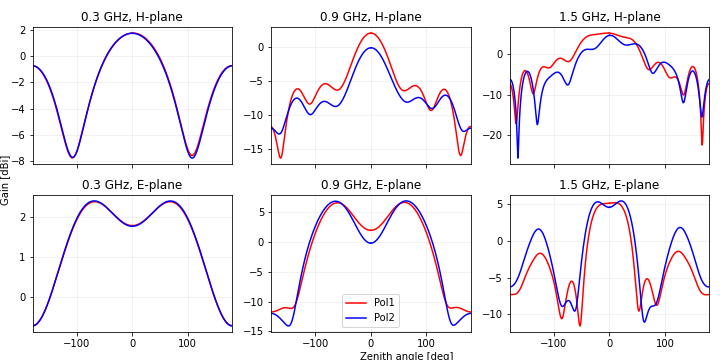}
    \caption{Simulations produced with CST studio suite (\citealt{CST2024}) of the beam for both polarizations of the CHORD feed. To compare with the feed + dish combination, see \citet{MacKay_2023}.}
    \label{fig:feedbeams}
\end{figure}

\appendix{}
\autoref{fig:feedbeams} shows the E- and H-plane CHORD feed beam slices for three frequencies. For the CHORD + dish beam see \citet{MacKay_2023}.

\FloatBarrier

\section*{Acknowledgments}
LB is supported by a Trottier Space Institute fellowship. IH is supported by Mitacs Accelerate Fellowship IT38715. MD is supported by a Canada Research Chair. AC, AWKL and JM are enabled by the Dunlap Institute for Astronomy and Astrophysics. The Dunlap Institute is funded through an endowment established by the David Dunlap family and the University of Toronto. 

%\printbibliography
\bibliographystyle{ws-jai}   
\bibliography{refs}

\end{document}